\documentclass[preprint,aps,floatfix,showpacs,preprintnumbers,amsmath,amssymb]{revtex4}
\usepackage{epsfig}
\usepackage{graphicx}
\usepackage{bbm}
\newcommand{\fr}{\frac}
\newcommand{\bea}{\begin{eqnarray}}
\newcommand{\eea}{\end{eqnarray}}
\begin{document}
\preprint{\vbox{\hbox{JLAB-THY-04-229} }}
\vspace{0.5cm}
\title{\phantom{x}
\vspace{0.5cm} $1/N_c$ Countings in Baryons}

\author{
J. L. Goity $^{a,b}$ \thanks{e-mail: goity@jlab.org}}

\affiliation{
$^a$ Department of Physics, Hampton University, Hampton, VA 23668, USA. \\
$^b$  Thomas Jefferson National Accelerator Facility, Newport News, VA 23606, USA. }
\date{\today}
\begin{abstract}
The $1/N_c$ power countings for baryon decays and configuration mixings are
determined by means of a non-relativistic quark picture. Such countings are expected
to be robust  under changes in the quark masses, and therefore valid as these become light. 
It is shown that excited baryons have natural  widths of  ${\cal{O}}(N_c^0)$. These dominant widths
are due to the decays that proceed directly to the ground state baryons, with cascade decays being suppressed
to ${\cal{O}}(1/N_c)$. Configuration mixings, defined as mixings between states belonging to different
$O(3)\times SU(2 N_f)$ multiplets, are shown to be sub-leading in an expansion in $1/\sqrt{N_c}$ when they involve the ground state baryons, while the mixings between excited states can be ${\cal{O}}(N_c^0)$.
\end{abstract}
\maketitle

\section{Introduction}

In this contribution to the proceedings  honoring  Professor Yuri Simonov on his  $70^{\rm th}$ birthday, I 
address some of the still open issues concerning the $1/N_c$ power counting for baryons. In particular, the power counting for the decay widths  of excited baryons as well as for configuration mixings  is analyzed. Several new conclusions result from this analysis. The $1/N_c$ expansion is one of the methods to which   Professor Simonov has made important contributions in his extensive work in QCD. 

The $1/N_c$ expansion was introduced by 't Hooft in a notable paper \cite{tHooft} thirty years ago. Although it has not lead to the ultimate goal of \lq\lq solving\rq\rq ~  QCD in its non-perturbative domain through  analytic tools, it has proven to be  powerful at the level of effective theory. The ability of implementing an ordering in powers of  $1/N_c$ at the hadronic level has lead to the understanding of numerous phenomenological facts. The large $N_c$ properties of mesons and their interactions can be established with little difficulty from the topological picture provided in 't Hooft's original paper. The implementation of the expansion in combination with Chiral Perturbation Theory (ChPT) in the light pseudoscalar sector \cite{Herrera} is one example of how it can be made to work at effective theory level. In the other qualitatively different sector, namely the baryons, the implementation of the $1/N_c$ expansion is substantially more involved. The pioneering work by Witten \cite{Witten} provided the guiding ideas for that implementation, and subsequent works by Gervais and Sakita \cite{Gervais} and by Dashen and Manohar \cite{Dashen} established the framework for the study of ground state baryons. 
 In this framework a  key role is played by the emergent spin-flavor symmetry  in the large $N_c$ limit. This provides the basis for the so called operator analysis that has been applied extensively to  the ground state baryons \cite{DJM,GSrefs}. The derivation of the spin-flavor structure can also be carried out in a less formal fashion than in \cite{Gervais,Dashen} by means of a non-relativistic quark picture \cite{March,Georgi1}.
At the effective theory level,  based on the results of the operator analysis, it has been possible to bring  the strictures of the $1/N_c$ expansion into baryon  ChPT  \cite{Flores}. Here, and because the baryon flavor multiplet contents  depend  on $N_c$, the formulation  of the effective theory is somewhat complicated. The operator analysis has been further extended to the sector of  excited baryons \cite{Georgi2,Goity,Pirjol}, where by now many results have been obtained \cite{CCGL,GSS,CC,GSS1,CoL3}. These results in particular  show that
the $1/N_c$ expansion can play an important and useful  role in sorting out the apparently complicated dynamics that determine the properties of baryonic resonances. 

Although the operator analysis for excited baryons is fairly well established, there have been a few open questions regarding the $1/N_c$ power counting for the decay widths, and the issue of configuration mixings, where states belonging to different spin-flavor and/or orbital multiplets, has been little studied. In this paper both aspects are addressed by means of a non-relativistic quark picture.  This picture is expected to reliably determine these  $1/N_c$ power countings.

\section{ Baryons in the large $N_c$ limit  }

 The first step towards implementing an analysis of baryons in the framework of the $1/N_c$ expansion is to establish the counting rules associated with the different operators that are needed in an effective theory. To proceed with this it is convenient to work in the limit where the quark masses are large enough for a non-relativistic picture to be reliable. In this way the problem of determining the $1/N_c$ counting is significantly simplified.  Because the $1/N_c$  counting should be largely  unaffected by the quark masses, the counting established in that limit should hold also for the situation where current quark masses are small. In the following, therefore, the discussion is based on such a  non-relativistic quark picture of baryons.

Using that approach, Witten \cite{Witten} showed that, obviously, baryon masses are proportional to $N_c$ while the baryon  size is only affected by corrections  ${\cal{O}}(1/N_c)$. In consequence, baryons are compact systems in large $N_c$, allowing for  the
rigorous  usage of the effective potential approach a la Hartree. A baryon state  can be expressed as follows:
\begin{equation}
\mid \Psi>=\frac{1}{N_c!}
\int \prod_{j=1}^{N_c} d^3 x_j\;
\Psi_{\xi_1,\cdots,\xi_{N_c}}(x_1,\cdots,x_{N_c})
\;\epsilon_{\alpha_1,\cdots,\alpha_{N_c}}
\;\mid x_1,\xi_1,\alpha_1;\cdots; x_{N_c},\xi_{N_c},\alpha_{N_c}\rangle,
\end{equation}
where $x_i$ are spatial positions, $\xi_i$ are spin-flavor indices and $\alpha_i$ are color indices. The states defined in terms of the non-relativistic quark creation operators with the standard anti-commutation relations are given by:
\begin{equation}
\mid x_1,\xi_1,\alpha_1;\cdots; x_{N-c},\xi_{N_c},\alpha_{N_c}\rangle=
\int \prod_{j=1}^{N_c} \frac{d^3k_j}{(2 \pi)^3}\;e^{i k_j x_j} \; 
q_{\xi_1 \alpha_1}^\dagger(k_1)\cdots q_{\xi_{N_c} \alpha_{N_c}}^\dagger
(k_{N_c})\mid 0\rangle .
\end{equation}
The wave functions $\Psi_{\xi_1,\cdots,\xi_{N_c}}(x_1,\cdots,x_{N_c})$ are totally symmetric under simultaneous permutations of positions and spin-flavor labels, and they satisfy the normalization:
\begin{equation}
\int \prod_{j=1}^{N_c} d^3x_j\;\Psi^*_{\xi_1,\cdots,\xi_{N_c}}(x_1,\cdots,x_{N_c})\;\Psi_{\xi_1,\cdots,\xi_{N_c}}(x_1,\cdots,x_{N_c})=1.
\end{equation}
A convenient basis of wave functions is furnished by functions factorized into a spatial and a spin flavor part added over permutations, namely,
\begin{equation}
\sum_{\sigma} \chi_{\xi_{\sigma_1}, \cdots, \xi_{\sigma_{N_c}}} \psi(x_{\sigma_1}, \cdots, x_{\sigma_{N_c}}). 
\end{equation}
In particular, it is convenient to take the spin-flavor wave functions $\chi$ to belong to an irreducible representation of the spin-flavor group $SU(2 N_f)$ if one is considering the case of $N_f$ flavors with degenerate or nearly degenerate masses. This means that such wave functions also belong to an irreducible representation of the permutation group of the $N_c$ indices. The crucial role played by  the spin-flavor group in the large $N_c$ limit makes this choice of basis natural. In a Hartree picture, the spatial wave function $\psi$ will have the form of a product of $N_c$  one-quark wave functions.
Ground state baryons in the large $N_c$ limit will therefore have wave functions of the form
\begin{equation}
\Psi^{GS}_{\xi_1,\cdots,\xi_{N_c}}(x_1,\cdots,x_{N_c})=
 \chi^S_{\xi_{1}, \cdots, \xi_{{N_c}}}\prod_{i=1}^{N_c}\phi(x_i),
\end{equation}
where the one-quark spatial wave function $\phi$ is an S-wave. Later on, the  admixture in the ground state of other spatial wave functions will be addressed (e.g. D-wave components) and  shown to be a sub-leading effect. The spin-flavor wave function must be  here totally symmetric, this being indicated by the upper label $S$. 

Excited baryons result from exciting one or  more quarks leaving a core of quarks in the ground state. A quark in the core has, up to corrections proportional to $1/N_c$, the same wave function as a  quark in the ground state baryons.  Although only  excited states with one excited quark are going to be discussed in detail, the generalization to two or more excited quarks can be carried out quite easily. The wave functions with one excited quark come in two types, namely  symmetric (S) and mixed-symmetric (MS) in spin-flavor. They are  respectively given by:
\begin{eqnarray}
\Psi^{S}_{\xi_1,\cdots,\xi_{N_c}}(x_1,\cdots,x_{N_c})&=&\frac{1}{\sqrt{N_c}}
\chi^S_{\xi_1, \cdots, \xi_{N_c}}
\sum_{i=1}^{N_c}\phi(x_1)\cdots \phi'(x_i)\cdots \phi(x_{N_c})\nonumber \\
 \Psi^{MS}_{\xi_1,\cdots,\xi_{N_c}}(x_1,\cdots,x_{N_c})&=&\frac{1}{\sqrt{N_c}(N_c-1)!}\sum_{\sigma}
\chi^{MS}_{\xi_{\sigma_1}, \cdots, \xi_{\sigma_{N_c}}}
\phi(x_{\sigma_{1}}) \cdots \phi(x_{\sigma_{N_c-1}})\phi'(x_{\sigma_{N_c}})
\end{eqnarray}
where $\phi'$ is the excited quark wave function which is taken to be orthonormal to the ground state wave function $\phi$. The mixed symmetry spin-flavor wave function $\chi^{MS}$ belongs to the representation with a Young tableaux having $N_c-1$ boxes in the first row and one box in the second row. In this case, the last index in the spin-flavor wave function  is the one associated with the excited quark.  The normalization of the spin-flavor wave functions is conveniently chosen to be unity so that the  one-quark spatial  wave functions have the same normalization. 

There is one point that needs mention. This is  the center of mass  degree of freedom that the wave functions used here  do not treat properly. The effects introduced by this defficiency are in general  subleading in   $1/N_c$ and  should not, therefore,   affect  the power countings addressed here. However, there  is the possibility that  countings, which  are suppressed  only on  orthogonality grounds of  one-quark wave functions used here, will be modified when the center of mass motion is properly treated. 

Since the contents of this paper have  to do with the $1/N_c$ counting of operator matrix elements, it is convenient at this point to define operators in the current framework. The non-relativistic quark field operator from which the various composite operators con be built reads:
\begin{equation}
q_{\xi \alpha}(x)=\int\frac{d^3 k}{(2 \pi)^3} \sum_{\lambda}\;  q_\lambda(k)\;  u^\lambda_{\xi \alpha}(k) \; e^{i k x}
\end{equation}
where $\lambda$ represents the polarization in color and spin-flavor. A natural choice for it is just in terms of the color and spin-flavor projections, that make the Pauli-spinor  to be delta functions, 
\begin{equation}
 u^{\lambda=\xi' \alpha'}_{\xi \alpha}(k)=\delta_{\xi\xi'} \delta_{\alpha\alpha'}.
\end{equation}
 Note that throughout the analysis that follows all operators have the same time argument, and therefore only the position vectors $x$ are displayed.

A color singlet 1-body local  operator  has the general form:
\begin{equation}
\Gamma_1(x)=q_{\xi \alpha}^\dagger(x) \;\Gamma_{\xi \xi'}(x) \; q_{\xi' \alpha}(x),
\end{equation}
where $\Gamma_{\xi \xi'}(x)$ is some functional operator acting on the quark fields. Note here that no generality for the application in this paper is gained by considering non-local operators.
An explicit evaluation of the matrix elements of 1-body  operators between generic baryonic states leads to:
\begin{eqnarray}
\langle \Psi'\mid \Gamma_1(x)\mid \Psi\rangle&=&
\int \prod_{j=1}^{N_c-1} d^3 x_j \;dx'_{N_c}dx_{N_c} \nonumber\\
&\times&
\Psi'^*_{\xi_1,\cdots,\xi_{N_c-1},\xi'_{N_c}}(x_1,\cdots,x_{N_c-1},x'_{N_c})\Gamma_{\xi' \xi}(x)
\Psi_{\xi_1,\cdots,\xi_{N_c-1},\xi_{N_c}}(x_1,\cdots,x_{N_c-1},x_{N_c})\nonumber\\
&\times& \int \frac{d^3k d^3k'}{(2 \pi)^6} \; e^{i k(x_{N_c}-x)}\; e^{-i k'(x'_{N_c}-x)}\; \delta_{\lambda (\xi_{N_c},\alpha)} \delta_{\lambda' (\xi'_{N_c},\alpha')} \;
u^{ \lambda'\dagger }_{\xi' \alpha'}(k')\; u^{ \lambda}_{\xi \alpha} (k).
\end{eqnarray}
 The sum over polarizations  boils down to a factor $N_c$ from the sum over color indices times  a spin-flavor factor  $\delta_{\xi \xi_{N_c}} \delta_{\xi' \xi'_{N_c}}$. This can be easily seen using the natural basis of Pauli spinors and performing the momentum integrations. As expected, the final form is:
\begin{eqnarray}
\langle \Psi'\mid \Gamma_1(x)\mid \Psi\rangle&=&N_c \int  \prod_{j=1}^{N_c-1} d^3 x_j \\
&\times& \Psi'^*_{\xi_1,\cdots,\xi_{N_c-1},\xi'}(x_1,\cdots,x_{N_c-1},x)\Gamma_{\xi' \xi}(x)
\Psi_{\xi_1,\cdots,\xi_{N_c-1},\xi}(x_1,\cdots,x_{N_c-1},x)\nonumber
\end{eqnarray}

As an illustration, consider the  important case  of the axial-vector current operator
\begin{equation}
A_{i a}(x)\equiv\frac{1}{4}\; q^\dagger_\alpha(x)\; \sigma_i t_a \; q_\alpha(x)=q^\dagger_\alpha(x)\; g_{i a}\; q_\alpha(x), 
\end{equation}
where $t_a$ are flavor generators. For the sake of simplicity, consider here the case of two flavors  and the matrix elements between ground state baryons. Applying  Equation (11), the matrix elements are given by:
\begin{equation}
\langle GS'\mid A_{ia}(x) \mid GS>= N_c \;
\chi'^{ S\dagger }_{\xi_1, \cdots, \xi_{N_c-1}, \xi'_{N_c}}\; (g_{ia})_{\xi'_{N_c}\xi_{N_c}}\;\chi^{ S }_{\xi_1, \cdots, \xi_{N_c-1}, \xi_{N_c}} \;\phi^*(x)\, \phi(x). 
\end{equation}
The matrix elements of the spin-flavor  generator 
$g_{ia}$ taken as shown in this equation are order $N_c^0$ when  the spin-flavor wave functions have  spin  ${\cal{O}}(N_c^0)$ (for two flavors, the symmetric spin-flavor states have all  $I=S$). Thus, the  result is  that the axial current matrix elements are order $N_c$. As a check, it is easy to  
 verify that the matrix elements of the spin and isospin operators are, as they should be,  ${\cal{O}}(N_c^0)$. Operators that, like the axial currents, receive the $N_c$ factor enhancement are called coherent operators. 

The above example leads to important implications.  Since  pions  couple to baryons through the axial-vector  current, the pion couplings are proportional to  $N_c/F_\pi={\cal{O}}(\sqrt{N_c})$ ($F_\pi$ scales as $\sqrt{N_c}$). As it is  briefly discussed in the next section, this large $N_c$ behavior
of the $\pi-$baryon couplings demands  the existence of a spin-flavor dynamical symmetry. Such a symmetry is  the  main reason why in the large $N_c$ limit there is a simplified picture of baryons.

Continuing with the issue of operators, consider now 2-body operators. A generic color singlet operator has the general form:
\begin{eqnarray}
\Gamma_2(x,y)=
q^\dagger(x) \otimes q(x) \; q^\dagger(y) \otimes q(y)\; \Gamma(x,y) 
\end{eqnarray}
where color and flavor indices are contracted through the tensor operator $\Gamma(x,y) $. A lengthier but equally straightforward evaluation as in the case of the 1-body operators gives the following expression  for 2-body operator matrix elements:
\begin{eqnarray}
\langle \Psi'\mid \Gamma_2(x,y)\mid \Psi \rangle&=&
\frac{N_c-1}{N_c}
\int \prod_{j=1}^{N_c-2} d^3 x_j \;dx_{N_c-1}dx_{N_c}dx'_{N_c-1} dx'_{N_c}\nonumber \\
&\times&
\Psi'^*_{\xi_1,\cdots,\xi_{N_c-2},\xi'_{N_c-1},\xi'_{N_c}}
(x_1,\cdots,x_{N_c-2},x'_{N_c-1},x'_{N_c})\nonumber\\
&\times&\Gamma(x,y)\;
\Psi_{\xi_1,\cdots,\xi_{N_c-2},\xi_{N_c-1},\xi_{N_c}}(x_1,\cdots,x_{N_c-2},x_{N_c-1},x_{N_c})\nonumber\\
&\times& 
\int 
\frac{d^3k_1 d^3k_2 d^3k'_1 d^3k'_2}{(2 \pi)^{12}} \;
e^{i (k_1(x_{N_c}-x)-k_1' (x'_{N_c}-x))}\;
 e^{i (k_2(x_{N_c-1}-y)-k_2' (x'_{N_c-1}-y))}\nonumber\\
&\times& 
u^{\dagger \lambda_1'}(k_1')\otimes  u^{ \lambda_1}(k_1)\;
 u^{\dagger \lambda_2'}(k_2')\otimes  u^{ \lambda_2}(k_2)\;
\delta_{\lambda_1, (\xi_{N_c}  \alpha_{N_c})} \delta_{\lambda_2, (\xi_{N_c-1}  \alpha_{N_c-1})}\nonumber\\
&\times& (
\delta_{\lambda'_1, (\xi'_{N_c}  \alpha_{N_c-1})}
\delta_{\lambda'_2,( \xi'_{N_c-1}  \alpha_{N_c})}- 
\delta_{\lambda'_1,( \xi'_{N_c}  \alpha_{N_c})}
\delta_{\lambda'_2,( \xi'_{N_c-1}  \alpha_{N_c-1})}).
\end{eqnarray}
This can be further evaluated leading to:
\begin{eqnarray}
\langle \Psi'\mid \Gamma_2(x,y)\mid \Psi \rangle&=&
\frac{N_c-1}{N_c}
\int \prod_{j=1}^{N_c-2} d^3 x_j \nonumber\\
&\times& \Psi'^*_{\xi_1,\cdots,\xi_{N_c-2},\xi'_{N_c-1},\xi'_{N_c}}
(x_1,\cdots,x_{N_c-2},x,y)\nonumber\\
&\times&(
\Gamma^{\xi'_{N_c} \alpha_{N_c-1},\xi'_{N_c-1} \alpha_{N_c}}_
{\xi_{N_c} \alpha_{N_c},\xi_{N_c-1} \alpha_{N_c-1}} (x,y)-\Gamma^{\xi'_{N_c} \alpha_{N_c},\xi'_{N_c-1} \alpha_{N_c-1}}_
{\xi_{N_c} \alpha_{N_c},\xi_{N_c-1} \alpha_{N_c-1}} (x,y))\nonumber\\
&\times&\Psi_{\xi_1,\cdots,\xi_{N_c-2},\xi_{N_c-1},\xi_{N_c}}(x_1,\cdots,x_{N_c-2},x,y).
\end{eqnarray}

An illustrative application of relevance for  baryon masses is the one-gluon exchange interaction. The 2-body operator associated with it is given by (disregarding spin-independent pieces that are order $1/m_q^2$ and other momentum dependent terms  which do not affect the point of the  discussion):
\begin{eqnarray}
{\cal{H}}_{OGE}(x-y)&\sim & g^2\; (-\frac{1}{\mid x-y\mid} \;q^\dagger(x)\frac{\lambda^A}{2} q(x)
\; q^\dagger(y)\frac{\lambda^A}{2} q(y) \nonumber \\
&+&\frac{1}{4 m_q^2}( (-4 \pi\; \delta^3(x-y)+\frac{1}{\mid x-y\mid^3})\; \delta_{ij}-3 \;\frac{(x-y)_i (x-y)_j}{\mid x-y\mid^5})
\nonumber\\
&\times& q^\dagger(x)\; \sigma_i \frac{\lambda^A}{2} q(x)\;
q^\dagger(y)\;\sigma_j \frac{\lambda^A}{2} q(y)),
\end{eqnarray}
where $\lambda^A$ are the $SU(N_c)$ generators in the fundamental representation. The first term in the right hand side is the color Coulomb interaction and the second term is the hyperfine interaction. Applying this to the ground state baryons the mass shift due to one-gluon exchange has the structure:
\begin{eqnarray}
\langle \Psi_{GS}\mid H_{OGE} \mid \Psi_{GS}\rangle & \sim & g^2\; \frac{N_c-1}{N_c} \;
\frac{(N_c^2-1)}{2} 
\int d^3x d^3y \;
\phi^*(x) \phi^*(y) \; \phi(x) \phi(y) 
\times 
(-\frac{1}{4 \mid x-y\mid}\nonumber \\ &+&
\frac{1}{4 m_q^2}\;
((-\pi \, \delta^3(x-y)+\frac{1}{4 \mid x-y\mid^3})\;\delta_{ij}
-\frac{3}{4}\;\frac{(x-y)_i\;(x-y)_j}{ \mid x-y\mid^5}  )\nonumber\\
&\times& \chi^{S*}_{\xi_1,\cdots,\xi'_{N_c-1},\xi'_{N_c}}\; s^i_{\xi'_{N_c-1} \xi_{N_c-1} }s^j_{\xi'_{N_c} \xi_{N_c} } \;
\chi^{S}_{\xi_1,\cdots,\xi_{N_c-1},\xi_{N_c}}).
\end{eqnarray}
The factor $(N_c^2-1)$ stems from the trace  over color indices. Taking into account that $g^2={\cal{O}}(1/N_c)$, the Coulomb interaction gives a contribution  ${\cal{O}}(N_c)$ that is independent of the spin-flavor of the state. 
The spin  matrix elements in the hyperfine term only contribute for $s_i s_j$ coupled to zero angular momentum. Thus,  for  states with spin ${\cal{O}}(N_c^0)$ the spin-flavor matrix elements in Equation (18) satisfy:
\begin{equation}
\langle \chi^S\mid s^i s^i \mid \chi^S\rangle= {\cal{O}}(\frac{1}{N_c} ) \; {\mathbf {1 }}+
{\cal{O}}(\frac{1}{N_c^2} ),
\end{equation}
i.e., they have a spin-flavor independent piece  ${\cal{O}}(1/N_c)$ and a spin-flavor dependent piece order ${\cal{O}}(1/N_c^2)$. This implies that the hyperfine interaction 
gives a spin-flavor independent mass shift of  ${\cal{O}}(N_c^0)$ and a breaking of spin-flavor symmetry of  
${\cal{O}}(1/N_c)$. This  important result establishes  that the spin-flavor tower of ground state baryons has splittings that are suppressed by $1/N_c$ for states with  spins ${\cal{O}}(N_c^0)$. 

Recently \cite{Manohar}, a  bosonic operator method has been introduced that should equally serve to carry out the derivations made in this section.

\section{ Ground State Baryons  }

The previous section gave the tools for determining the counting in the $1/N_c$ expansion associated with various  matrix elements. In all cases  the counting is in the end determined by a few characteristics of the operator  being  considered, namely their  n-bodyness and  spin-flavor structure, and as shown later,  by the spin-flavor representation to which the states belong and the degree of excitation of the states (number of excited quarks).  This permits the  implemention of  the counting at the   effective theory level. This section briefly outlines how this has been carried out for the ground state baryons. 

The result at the end of the previous section can be put in a more general framework in which
the constraints  of unitarity in pion-baryon scattering demand   a dynamical  spin-flavor symmetry \cite{Gervais,Dashen}.   This symmetry is of course satisfied by the non-relativistic quark picture. Thanks to the spin-flavor symmetry, ground state baryons can be chosen to fill an $SU(2 N_f)$ multiplet, namely, the totally symmetric irreducible representation with $N_c$ spin-flavor indices.  Any color singlet operator 
in QCD will then  be represented at the level of the effective theory by a series of composite effective operators 
ordered  in powers of $1/N_c$. These composite operators, 
when acting on a specific spin-flavor representation,
can be  further  represented via the Wigner-Eckart theorem by   appropriate products of  generators of the spin-flavor group  \cite{Dashen,DJM}. For instance,  the matrix elements of the QCD Hamiltonian between ground state baryons give the masses of these states. The most general mass operators that one could write down  are proportional to: $\mathbf{1}$, $S^2$, $G^2$, $T^2$, etc. Here $S_i$,
$T_a$ and $G_{ia}$ are the generators of $SU(2 N_f)$, which in the non-relativistic quark picture are given by:
\begin{eqnarray}
S_i&=&\frac{1}{2}\;\int  q^\dagger(x)\; \sigma_i \;q(x)\; d^3x\nonumber\\
T_a&=&\frac{1}{2}\;\int  q^\dagger(x)\; t_a \;q(x)\; d^3x\nonumber\\
G_{ia}&=&\frac{1}{4}\;\int  q^\dagger(x)\; \sigma_i t_a \;q(x)\; d^3x.
\end{eqnarray}
 The $1/N_c$ counting associated with an n-body effective operator is given by the general formula:
\begin{equation}
\nu=N_c^{(1-n)}\times N_c^\kappa,
\end{equation}
where $\nu$ is the order in $1/N_c$ of the matrix elements of the operator. The first factor in the right hand side  results from the fact that in order to generate  an effective n-body operator starting from a 1-body operator at the QCD level,  $n-1$ gluon exchanges are necessary (this factor is usually included in the definition of the effective operator as shown below), and the second factor results from the number $\kappa$ of coherent factors (the generator $G$ above is a coherent factor as the result in 
the previous section about the axial current matrix elements show). As illustration, consider the mass operator for the ground state baryons (for the sake of simplicity take two degenerate flavors where $S=I$). The most general mass operator one can write down is therefore:
\begin{equation}
 H_{GS~mass}=N_c\; m_0 \;{\mathbf {1 }} + m_1 \frac{1}{N_c} S^2+ m_1'  \frac{1}{N_c}\; G^2+{\rm 3\!-\!\! body}+\cdots~.
\end{equation}
The first term gives the overall spin-flavor singlet ${\cal{O}}(N_c)$ mass, the second term gives the  ${\cal{O}}(1/N_c)$ mass splittings. On the other hand, the third term, which according to the counting rule given above is  ${\cal{O}}(N_c)$, turns out to be linearly dependent up to  ${\cal{O}}(1/N_c)$ with the other two operators, and therefore to that order it can be eliminated (a series of such reduction rules have been established \cite{DJM}). Thus, up to and including ${\cal{O}}(1/N_c)$ effects the GS baryon masses can be represented by the first two terms on the right hand side of Equation (22). 
GS matrix elements associated with other operators (axial currents, magnetic moments, etc.) have been extensively analyzed elsewhere \cite{GSrefs}.

\section{ Excited  Baryons }

The existence of a spin-flavor symmetry at the level of GS baryons suggests that such a symmetry ought to play an important role in excited baryons.  An approach that has been proposed \cite{Georgi1,Goity,Pirjol}, which is the natural one in the non-relativistic quark picture, is to describe the excited baryons using a   basis of states filling  multiplets of the $O(3)\times SU(2 N_f)$ group. The $O(3)$ group has as generators the orbital angular momentum operators. While in the GS baryons the spin-flavor symmetry is broken at  ${\cal{O}}(1/N_c)$, in the excited baryons the extended  $O(3)\times SU(2 N_f)$ symmetry can be broken at zeroth order \cite{Goity}. The reason for this zeroth order breaking is the possibility of spin-orbit couplings.  In the quark picture this can be easily demonstrated.  The induced Thomas precession  term, which is represented by a 1-body  operator of  ${\cal{O}}(N_c^0)$, reads
\begin{equation}
H_{SO}= w\; \ell \cdot S  ~,
\end{equation}
where the parameter $w$ contains the details about the binding of the excited quark in the baryon and $s$ is the spin operator. Calculating its matrix elements for excited states with the generic wave functions
\begin{equation}
\Psi'_{\xi_1,\cdots,\xi_{N_c}}(x_1,\cdots,x_{N_c})=
\frac{1}{\sqrt{N_c}(N_c-1)!}\sum_{\sigma}\;
\chi'_{\xi_{\sigma_1}, \cdots, \xi_{\sigma_{N_c}}}
\phi(x_{\sigma_{1}}) \cdots \phi(x_{\sigma_{N_c-1}})\phi'(x_{\sigma_{N_c}}),
\end{equation}
the spin-orbit mass shift has the form:
\begin{equation}
\langle \Psi' \mid H_{SO} \mid \Psi'\rangle=w\;\ell_i\; \langle \chi\mid  s_i \mid\chi \rangle,
\end{equation}
where in the  spin-flavor  matrix element the spin operator $s_i$ acts only on the spin of the excited quark (i.e., the last index of the spin-flavor wave function). If $\chi$ belongs to the symmetric spin-flavor representation,  $\langle \chi\mid  s_i \mid\chi \rangle$ is  ${\cal{O}}(1/N_c)$, while  if   it belongs to the MS representation the result is  
${\cal{O}}(N_c^0)$. Thus, the spin-orbit coupling affects states in the MS representation at order $N_c^0$. Among other effects, this leads to a breaking of spin-flavor symmetry at the same order \cite{Goity}. This would seem to have bad consequences for the spin-flavor symmetry in  MS states, but it turns out not to be so. First, the spin-orbit breaking leaves a remnant
symmetry associated with states of the core of $N_c-1$ quarks as shown in \cite{PirjolSchat,CoL1}. This remnant symmetry is broken at sub-leading order by hyperfine effects. Second, as various detailed analyses have shown \cite{Goity,CCGL,GSS}, the spin-orbit effects in  the $SU(6)$ 70-plet of negative parity baryons are  unnaturally small for  not as yet  fully understood dynamical reasons (substantially smaller than the sub-leading hyperfine effects). From a practical point of view, this implies that the  basis of states in terms of multiplets of  $O(3)\times SU(2 N_f)$ is  very useful. Other operators that couple the orbital angular momentum do contribute to zeroth order spin-flavor breaking. The complete analysis of the negative parity baryon masses
\cite{Goity,CCGL,GSS} shows in general that the zeroth order breaking is unnaturally small.

The operator analysis in the case of excited baryons proceeds in analogy with that for the GS baryons, except that now one has an extended set of generators that includes the orbital angular momentum generators. The details of the procedure have been given elsewhere \cite{Goity,CCGL,GSS} and will not be repeated here.

The main point of this paper  is to establish a few  results of general validity and importance for excited baryons. These have to do with the $1/N_c$ counting for the various decays of excited baryons, and with  the possible mixings of  $O(3)\times SU(2 N_f)$ multiplets (configuration mixings).

\subsection{Decays}

 The original  work of Witten indicated, correctly as shown below, that excited baryons have widths order $N_c^0$. This is in sharp contrast with mesons, which become stable in the large $N_c$ limit, their widths being  ${\cal{O}}(1/N_c)$ or higher. More recently, some questions have been raised about the general validity of that zeroth order result. An appraisal of the situation can be found in Ref. \cite{CoL2}.

Since the determination of the $1/N_c$ counting for the widths will not depend on fine details, a model for the  decay process through one-pion emission  can be used. The emission of an $\eta$ meson such as in the decay $N^*(1535)\to N\eta$ will be briefly mentioned as well. The model for the discussion is the chiral quark model \cite{ChQM} in which the pion couples to the constituent quark according to
\begin{equation}
H_{ChQM}=-\frac{g_A^q}{ 4 F_\pi}\int d^3 x\; \partial_i \pi_a (x)\; q^\dagger(x)\; \sigma_i \tau_a \;q(x), 
\end{equation}
where $g_A^q$ is the constituent quark axial-vector coupling of order $N_c^0$. Using the 1-body  Equation (11),  the amplitudes for the various possible transitions are readily calculated. 
The first type of transitions are the ones that occur   within a multiplet. The dominant amplitude for these transitions is easy to obtain using the result already derived  in section II for the matrix elements of the axial-vector currents  to which the pion couples as shown by the above Hamiltonian. For GS as well as excited baryons these  amplitudes are in general proportional to $\sqrt{N_c}$. The GS  widths in this case are suppressed by phase space as the baryon states for which the amplitude is order $\sqrt{N_c}$  have mass differences that are order $1/N_c$.  Because  these transitions are 
 P-wave, the end result is that the partial widths are   ${\cal{O}}(1/N_c^2)$.  One such a transition is the $\Delta\to \pi N$ transition. Note that for states at the top of the GS tower the amplitude is proportional to $1/\sqrt{N_c}$, and since the splittings are zeroth order, the widths are ${\cal{O}}(1/N_c)$. In the real world the $\Delta$ is the second and top state of the tower, and its width should therefore be  between the two limits.  The transitions within an excited multiplet containing  zeroth order mass splittings can be shown to have amplitudes
$\sim\sqrt{N_c}$ only between  states whose relative mass splittings are ${\cal{O}}(1/N_c)$. The reason is that these amplitudes   only change the core's state, and such a change can only affect the energy level through a 2-body mass operator. Since the
core piece of  matrix elements of the 2-body operator can only be affected at ${\cal{O}}(N_c^0)$  by that  change of the core state, the effect on the energy level must be at most ${\cal{O}}(1/N_c)$.

 The second kind  of  transitions are from excited baryons to GS baryons. The decay  amplitudes in this case have  the form
\begin{equation}
\langle\Psi_{GS}+\pi\mid\Psi'\rangle=\frac{g_A^q}{ F_\pi}\; \sqrt{N_c}\; k_{\pi_i}
\int d^3x\; e^{ik_\pi x} \; \phi^*(x) \phi'(x) \; \langle \chi^S\mid g_{ia} \mid \chi'\rangle, 
\end{equation}
where
\begin{equation}
 \langle \chi^S\mid g_{ia} \mid \chi'\rangle=\chi^{S*}_{\xi_1\cdots\xi'_{N_c}}
(g_{ia})_{\xi'_{N_c}\xi_{N_c}}\; \chi'_{\xi_1\cdots\xi_{N_c}}.
\end{equation}
These spin-flavor matrix elements are order $N_c^0$ for both S and MS   $\chi'$. Thus, irrespective of the spin-flavor character of the excited state, the decay rate to the GS via one-pion emission is of zeroth order in the $1/N_c$ expansion. 

The third type of transitions are those between different excited baryon multiplets, where now the amplitude becomes
\begin{equation}
\langle\Psi''+\pi\mid\Psi'\rangle=\frac{g_A^q}{ F_\pi} \; k_{\pi_i}\;
\int d^3x \;e^{ik_\pi x}\; \phi''^{*}(x) \phi'(x)\; \langle \chi''\mid\; g_{ia}\; \mid \chi'\rangle.
\end{equation}
This amplitude is similar to that for the transition to the ground state except for the absence of the factor $\sqrt{N_c}$. Thus, the transition amplitudes between excited states are generically suppressed by  a factor $1/\sqrt{N_c}$ with respect to the ones to the GS.  This has an important implication, namely, in large $N_c$ limit the dominant channel of decay for excited baryons is the direct decay to the ground state.   If $\eta$ emission is considered, the analysis is similar except that $g_{ia}$ is replaced by $s_i$. The result in this case is that the corresponding spin-flavor matrix elements are ${\cal{O}}(N_c^0)$ if $\chi'$ is in the MS representation and  ${\cal{O}}(1/N_c)$ if it is in the symmetric representation. These countings have  been used to implement the operator analysis for the decays of the negative parity baryons \cite{Georgi2,GSS2} as well as the Roper multiplet \cite{CC}.

The discussion above is valid for excited states where only one quark is excited. If more than one quark is excited further suppressions occur, namely a factor $1/\sqrt{N_c}$ per excited quark. In these cases it is necessary to consider
  2- or higher-body decay operators. Leaving the details out, for two excited quarks the decay amplitude with emission of a single pion is order $1/\sqrt{N_c}$ for decay to GS and order $1/N_c$ for decay to other excited states. This implies that the decay rate via single pion emission is order $1/N_c$. The total width is  however expected to be of zeroth order. 
The answer to this riddle seems to be  that the two-pion emission is order $N_c^0$.
One way to see this is shown by the diagrams in  the figure, where the suppression factor
in one amplitude, the one connecting excited and ground states, is compensated by the enhancement factor of the amplitude connecting states in the same multiplet. This implies that these excited states decay predominantly by emitting two pions. There is  a subtlety to be dealt with here. It has to do with the  application of the same mechanism to the baryons with only one excited quark. Naively, for these states the dominant contribution for two-pion emission  would then be order $\sqrt{N_c}$, which cannot be right. Such an unacceptable  contribution must be  cancelled  through the interference of the  various   baryonic intermediate states. This results from   consistency relations similar to those that eliminate the order $N_c$ terms in  $\pi-$baryon scattering \cite{Dashen,Pirjol}. This issue of consistency relations involving excited baryons in general has not been extensively analyzed,  and it certainly deserves further consideration. 

The discussions of the decays have been carried out in a limited framework. This, however, should fully clarify the picture:
for excited baryons the decay amplitudes are of zeroth order, so that when building an effective theory such as in \cite{Georgi2,CC,GSS2}, it is necessary to carefully trace the power counting, namely, for one-pion decays there is a factor $1/F_{\pi}$ whose origin is rather obvious, and the less obvious factor of $\sqrt{N_c}$ that is shown in Equation (27) whose origin
is very clear in the model considered here.

\begin{figure}
\vspace{3cm}
\centerline{\epsfig{file=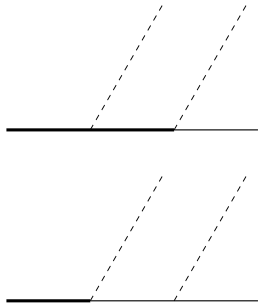,height=4.cm,width=3.4cm}}
\caption{The thick solid line represents  excited baryons belonging to a single multiplet, the thin one represents  a ground state baryon, and the dashed lines represent pions. The vertices connecting an excited and ground state baryon are proportional to  $1/\sqrt{N_c}$ for two-quark excited baryons,  while the other vertices are proportional to $\sqrt{N_c}$.}
\end{figure}

\subsection{Configuration mixing}
All large $N_c$  analyses carried out  in the literature  disregard the mixing
of different spin-flavor as well as orbital states. As the following discussion shows, this is in most cases the correct thing to do. The only relevant mixings that are of any significance  are mixings between states belonging to different representations of   $O(3)\times SU(2 N_f)$. For instance, the admixture of $\ell=2$ states in the ground state baryons, and/or the admixture of states belonging to the S and MS spin-flavor representations.  In this subsection the $1/N_c$ countings for the various mixings are obtained. 

The Hamiltonian  that drives the mixings is    rotationally invariant, which means that  under $O(3)\times SU_{spin}(2)$ it  must transform, in obvious notation,  as $(\ell, s=\ell)$. It is also taken to be flavor symmetric.

There is only one 1-body operator that can produce configuration mixing, namely the spin-orbit operator \cite{CoL2}. This operator can only give $\Delta\ell=0$ mixings of S and MS spin-flavor representations. The mixing amplitudes are given by a formula similar to (25):
\begin{equation}
\langle \Psi''\mid H_{SO}^{mix} \mid \Psi'\rangle= c\;\ell_i\; \langle \chi^{MS}\mid  s_i \mid\chi^S \rangle,
\end{equation}
where $c$ is of zeroth order and the spin-flavor matrix element is  ${\cal{O}}(N_c^0)$. Thus, at the 1-body level there is zeroth order  $\Delta\ell=0$ mixing between states with $\ell>0$.

At the level of  2-body operators other mixing possibilities exist, in particular mixings   involving the GS baryons. A generic 2-body  Hamiltonian that contributes to mixing is:
\begin{equation}
 H_{mix}=\frac{1}{N_c} \int d^3x\; d^3 y \; L_{ij}(x,y) \;{\cal{S}}_{ij},
\end{equation}
where  $L_{ij}$ is a tensor operator of up to rank 2, and ${\cal{S}}_{ij}$ is a spin-flavor operator that is a flavor singlet.

The mixing amplitude for two generic states is readily determined by applying Equation (16). Up to sub-leading terms in $1/N_c$ the result is:
\begin{eqnarray}
\langle\Psi'\mid H_{mix} \mid \Psi \rangle&=& - N_c \int \prod_{j=1}^{N_c}
d^3 x_j\;
\Psi'^*_{\xi_1\cdots\xi'_{N_c-1}\xi'_{N_c}}(x_1,\cdots,x_{N_c-1},x_{N_c})\nonumber\\
&\times&L_{ij}(x_{N_c-1},x_{N_c})\;
\Psi_{\xi_1\cdots\xi_{N_c-1}\xi_{N_c}}(x_1,\cdots,x_{N_c-1},x_{N_c})\nonumber\\
&\times&({\cal{S}}_{ij})^{\xi'_{N_c-1},\xi'_{N_c}}_{\xi_{N_c-1},\xi_{N_c}}.
\end{eqnarray}
There are several cases to be considered.  The  first case is   configuration mixings involving the GS baryons. The mixing amplitude  with excited states having  wave functions  of the form given by Equation (24)  becomes
\begin{eqnarray}
\langle\Psi'\mid H_{mix} \mid \Psi_{GS}\rangle&=&-\sqrt{N_c}\int d^3x\; d^3y\; (\phi^*(x)\phi'^*(y)+x\leftrightarrow y)\nonumber\\
&\times& L_{ij}(x,y)\; \phi(x)\phi(y)\; \langle \chi'\mid\; {\cal{S}}_{ij}\;\mid\chi\rangle,
\end{eqnarray}
where the 2-body spin-flavor matrix element is defined by
\begin{equation}
\langle \chi' \mid\; {\cal{S}}_{ij}\; \mid\chi\rangle=
\chi'^{*}_{\xi_{1},\cdots,\xi'_{N_c-1},\xi'_{N_c}}\;\;
({\cal{S}}_{ij})^{\xi'_{N_c-1},\xi'_{N_c}}_{\xi_{N_c-1},\xi_{N_c}}\;\;
\chi_{\xi_{1},\cdots,\xi_{N_c-1},\xi_{N_c}}.
\end{equation}
The order in $1/N_c$ of the amplitude is determined by this latter matrix element. There are a couple of cases to be considered. One case is when 
$\chi'$ is in the symmetric representation, implying that the excited state must have $\ell\neq 0$ for an observable configuration mixing to take place. Since the two body spin-flavor operator can be  at most of rank 2, parity conservation  implies that $\ell=2$. The second case is when $\chi'$ is in the MS representation, where now both possibilities exist, namely $\ell=0$ and 2.
The most general forms of ${\cal{S}}_{ij}$ are ${\cal{S}}_{ij}=s_i \otimes s_j$
and ${\cal{S}}_{ij}=g_{ia} \otimes g_{ja}$ coupled to $\ell=0$ and 2.
The  countings in $1/N_c$ of the  various relevant matrix  elements of these operators can be obtained by explicit evaluation and are shown in the table.
\begin{table}[]
\vspace{2cm}
\begin{tabular}{lcccc}
\hline
\hline
Matrix elements &{$\ell=0$}&{$\ell=2$} \\
\hline
\hline
$\langle S\mid s_i\otimes s_j\mid S\rangle$ & * ${\cal{O}}( 1/N_c) {\mathbf {1 }}$  & ${\cal{O}}(1/ N_c^2)$  \\
$\langle S\mid g_{ia} \otimes g_{ja}\mid S\rangle$ & * ${\cal{O}}( N_c^0) {\mathbf {1 }}+{\cal{O}}( 1/N_c^2)$ & ${\cal{O}}( 1/N_c^2)$  \\
$\langle MS\mid   s_i\otimes s_j \mid S\rangle $ & ${\cal{O}}( 1/N_c) $  & ${\cal{O}}( 1/N_c)$  \\
$\langle MS\mid g_{ia}\otimes g_{ja}\mid S\rangle $ & ${\cal{O}}( 1/N_c^2)$& ${\cal{O}}( 1/N_c)$  \\
$\langle MS\mid s_i\otimes s_j\mid MS\rangle $ & * ${\cal{O}}( 1/N_c)  $   & ${\cal{O}}( 1/N_c)$  \\
$\langle MS\mid g_{ia}\otimes g_{ja}\mid MS\rangle$&~~~~* ${\cal{O}}( N_c^0)  {\mathbf {1 }}+{\cal{O}}( 1/N_c^2)$ ~~~~ & ${\cal{O}}( N_c^0) $ \\
\hline \hline
\end{tabular}
\caption{List of spin-flavor matrix elements relevant to configuration mixings and their counting in $1/N_c$. Here, {\bf 1 } denotes the singlet spin-flavor operator. The * indicates entries that produce irrelevant configuration mixings. }
\label{operators}
\vspace{4cm}
\end{table}
These countings  imply that the mixing in the GS baryons are as follows: mixings with states in the symmetric spin-flavor representation (which, as mentioned earlier, require the spin-flavor operator to be $\ell=2$) are order $1/N_c^{3/2}$, while mixings with a MS representation are in general order $1/\sqrt{N_c}$. Thus, the mixing effects affect ground state baryons primarily at the level of their spin-flavor representation content. Notice that these mixings can only  affect the mass splittings at  ${\cal{O}}(1/N_c)$ as it should be.

Finally the configuration mixings between excited states are given by generic matrix elements:
\begin{eqnarray}
\langle\Psi''\mid\; H_{mix}\; \mid \Psi'\rangle&=&-\int d^3x\; d^3y\; (\phi^*(x)\phi''^{*}(y)+ x\leftrightarrow y)
\nonumber\\
&\times& L_{ij}(x,y)\; \phi(x)\phi'(y)\; \langle \chi''\mid\; {\cal{S}}_{ij}\;\mid\chi'\rangle.
\end{eqnarray}
 Using the countings of the table, if $\chi'$ and $\chi''$ are both symmetric the configuration mixing is order
$1/N_c^2$, and order $1/N_c$ if only  one of them is symmetric. Note here that the effective operator 
$1/N_c/;\ell^{(2)}\; g\; G^c$,  unlike the 1-body spin-orbit operator,  gives sub-leading mixings. If both states  are in the MS representation, observable configuration mixing only occurs
for $\ell=2$, and as the  last entry in the table shows, this is order $N_c^0$. Thus, with the exception of the latter case, configuration mixings of excited states induced by 2-body operators are suppressed by $1/N_c$. 

From the discussion above, the zeroth order mixings affecting excited states come in two types: $\Delta \ell=0$ mixings that require states with $\ell>0$ and which mix S and MS representations, and $\Delta\ell=2$ mixings involving only MS states.  The strength of zeroth order mixings depends on the dynamics. Because this type of mixing requires spin and orbital couplings, the observed weakness of the spin-orbit coupling that consistently  results from analyses of the baryon spectrum hints that the mixing effect is, for little understood reasons, dynamically suppressed. An example of zeroth order configuration mixing would be   $\ell=3$ components in the negative parity SU(6) 70-plet wave functions, or the mixing between an $\ell=1$ 56-plet with the $\ell=1$ 70-plet as noted in \cite{CoL2}. Finally, it is not difficult to extend the discussion to  configuration mixings involving states with more than one excited quark. Such mixings are  suppressed by extra factors of $1/\sqrt{N_c}$.

\section{Conclusions}

The $1/N_c$ counting in baryons discussed in this paper from the point of view of a non-relativistic quark picture 
gives a good perspective about what the physics of baryons would be like in a world with a large number of colors.
It is expected that the countings obtained here will hold also as the quark masses become small.

The results show  that excited baryons have finite widths in that world. This also means that a picture in which excited baryons are resonances of pions and ground state baryons is perfectly viable \cite{CoL3,Lutz}. The other observation is that
the dominant decay channel is always the one that leads most directly to the ground state; cascading through
other excited states belonging to different multiplets  is a sub-leading effect. 

Configuration mixing is an issue that will need further understanding because it can occur in some cases at zeroth order in $1/N_c$. This posses some difficulties of principle for the study of the excited baryon sector. The mixings affecting the ground state baryons are suppressed at least by one factor $1/\sqrt{N_c}$. The dominant configuration  mixing in this case involves  spin-flavor mixing.
On the other hand, the mixings of excited multiplets can be  ${\cal{O}}{N_c^0}$.  Because these zeroth order mixings are driven by the couplings of the orbital angular momentum, and  orbital couplings are phenomenologically known to be small, it is very plausible that the mixings  are dynamically suppressed. This is an open issue, however, which deserves further scrutiny.

It is not quite clear how well or how poorly  the general structures implied by the countings just discussed  survive in the real world with $N_c=3$. This is a difficult issue involving the convergence of the expansion that is far from being established. Further phenomenological analyses
such as those carried out for masses, decays and scattering, and  the applications of the expansion to lattice QCD
 simulations of excited baryon will eventually clarify the issue.

\acknowledgements

I would like to thank Professor  Yuri Simonov for enlightening discussions about the $1/N$ expansion  during his visit to Jefferson Lab. I would also like
to thank  Winston Roberts for pointing out the importance of the two-pion decay channels,  Carlos Schat and Norberto Scoccola for very   useful discussions on the problems  addressed in   this paper, and Richard Lebed for a discussion that served to expose a mistake in the first version of this paper.
This work was partially supported by the National Science Foundation  through grants
PHY-9733343 and PHY-0300185, and  by  Department of Energy  contract DOE-AC05-84ER40150.

\end{document}